\documentclass[fleqn,usenatbib]{mnras}

\usepackage{newtxtext,newtxmath}
\usepackage[T1]{fontenc}
\usepackage{ae,aecompl}

\usepackage{graphicx}	% Including figure files
\usepackage{amsmath}	% Advanced maths commands
\usepackage{amssymb}	% Extra maths symbols

\title[Bar rejuvenation in S0 galaxies?]{Bar rejuvenation in S0 galaxies?}

\author[Barway \& Saha]{
Sudhanshu Barway,$^{1}$\thanks{E-mail: sudhanshu.barway@iiap.res.in (SB)}
and Kanak Saha$^{2}$
\\
$^{1}$Indian Institute of Astrophysics (IIA), II Block, Koramangala, Bengaluru 560 034, India\\
$^{2}$Inter-University Centre for Astronomy and Astrophysics, Pune 411007, India}

\date{Accepted XXX. Received YYY; in original form ZZZ}

\pubyear{2020}

\begin{document}
\label{firstpage}
\pagerange{\pageref{firstpage}--\pageref{lastpage}}
\maketitle

\begin{abstract}
Based on the colour measurements from a multi-band, multi-component 2D decomposition's of S0 and spiral galaxies using SDSS images, we found that bars are bluer in S0 galaxies compared to the spiral galaxies. Most of the S0s in our sample have stellar masses $\sim L_{*}$ galaxies. The environment might have played an important role as most of the S0s with bluer bars are in the intermediate-density environment.  The possibility of minor mergers and tidal interactions which occurs frequently in the intermediate-density environment might have caused either a bar to form and/or induce star formation in the barred region of S0 galaxies. The underlying discs show the usual behaviour being redder in S0s compared to spiral galaxies while the bulges are red and old for both S0 and spiral galaxies. The finding of bluer bars in S0 galaxies is a puzzling issue and poses an interesting question at numerical and theoretical studies most of which shows that the bars are long-lived structures with old stellar populations.
\end{abstract}


%%%%%%%%%%%%%%%%% BODY OF PAPER %%%%%%%%%%%%%%%%%%
\section{Introduction} 
\label{sec:intro}
In the hierarchical structure formation scenario \citep{whiterees1978}, disc galaxies are believed to have formed via mergers of smaller components. Local disc galaxies are often found to have more than one structural sub-components such as bulges, bars, lenses, rings, see \cite{KK2004} for a comprehensive review. In fact, more than  60\% of disc galaxies that dominate the star formation in the local universe are known to host strong bars \citep{Eskridgeetal2000,MenendezDelmestreetal2007,Barazzaetal2008,Aguerrietal2009}. Although bars exist at redshift $z \sim 1 -  2$, albeit with a lowered frequency \citep{Elmegreenetal2004,Jogeeetal2004,shethetal2008}, it remains unclear when disc galaxies formed their first bar. Some orbital studies and numerical simulations have shown that bars could be destroyed as a result of growing central mass concentration or growing supermassive blackholes or as a result of cold gas inflow to the central region \citep{PfennigerNorman1990,ShenSellwood2004,HozumiHernquist2005, Bournaudetal2005}  while some simulations indicate that bars are robust and once formed, it is hard to dissolve them \citep{Athanassoulaetal2005,Kraljicetal2012}. So in the absence of legitimate evidences and clues, it remains a harder problem to understand whether the same bar continued to exist in the present day-galaxies or it has been dissolved and rejuvenated \citep{BournaudCombes2002}.

One of the possible ways to resolve this issue would be to investigate the stellar population in the bar region and compare with the rest of the galaxy.  This is a challenging task to accomplish as stars in the bar, bulge and the disc are often mixed and we only see what is along the line of sight. The bar, being one of the strongest non-axisymmetric structure in the disc, plays an active role in the mixing of stars in the galaxy \citep{Fraser-McKelvieetal2019}. Not only that, it is the driver of gas inflow in the central region and thereby the cause of star formation activity \citep{Aguerri1999} that is seen in the inner bar region, especially around the inner Lindblad resonance (ILR) in the form of a star-forming ring e.g., in NGC 1097 \citep{Martinetal2015,Prietoetal2019}. Nevertheless, when looked through the infrared, most bars in the local universe to a large extent are composed of a red and old stellar population \citep{Perezetal2009,Sanchez-Blazquezetal2011}, except probably at the resonance locations \citep{Wozniak2007}. To understand the finer details on the spatial variation of stellar population and star formation histories, one would require high spatial resolution IFU data on a statistically significant sample of disc galaxies. In absence of that, stellar population studies of galaxies beyond the Local Group have been investigated primarily using the broad-band colours. Although in the literature \citep[\& references therein]{Barway.et.al.2013}, the observed stellar population has been studied using a galaxy's global colour, stellar population properties are known to vary between the the different structural components of a disc galaxy e.g., its bulge, disc or in some cases the bar \citep{Head.et.al.2014, Sanchez-Blazquezetal2011,Hudson.et.al.2010, Simard.et.al.2011, Gadotti2009}. The multi-band decomposition of a barred galaxy into bulge, bar and disc would thus allow one to study the distribution of colours for each component and its stellar population properties within limitation. 
%
%
%%%%%%%%%%%%%%%%%%%%%%%%%%%%%%%%%%%%%%%%%%%%%%%%%%%%%%%%%%%%%%%%%%%%%%
\begin{figure*}
\includegraphics[scale=0.42]{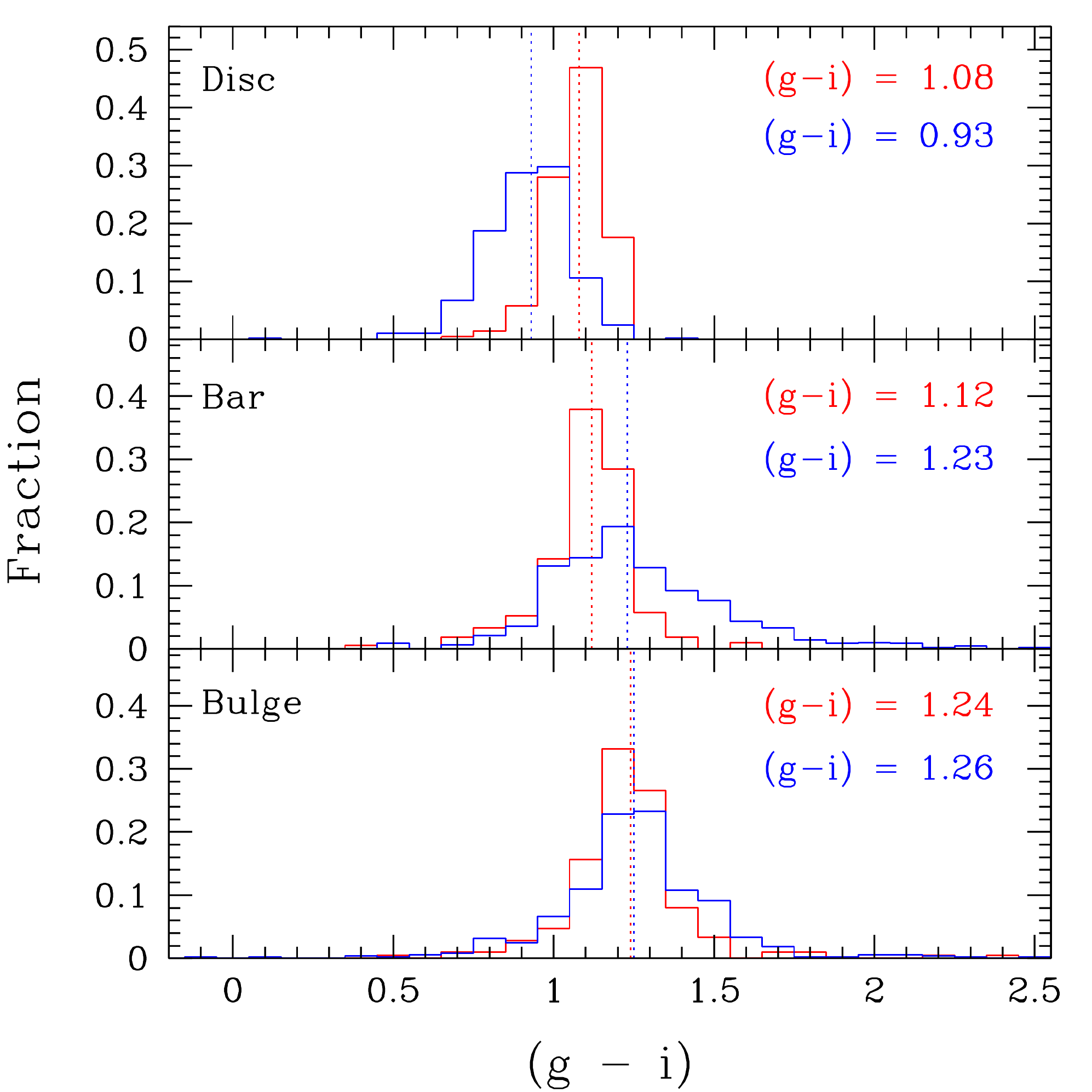}%
\includegraphics[scale=0.42]{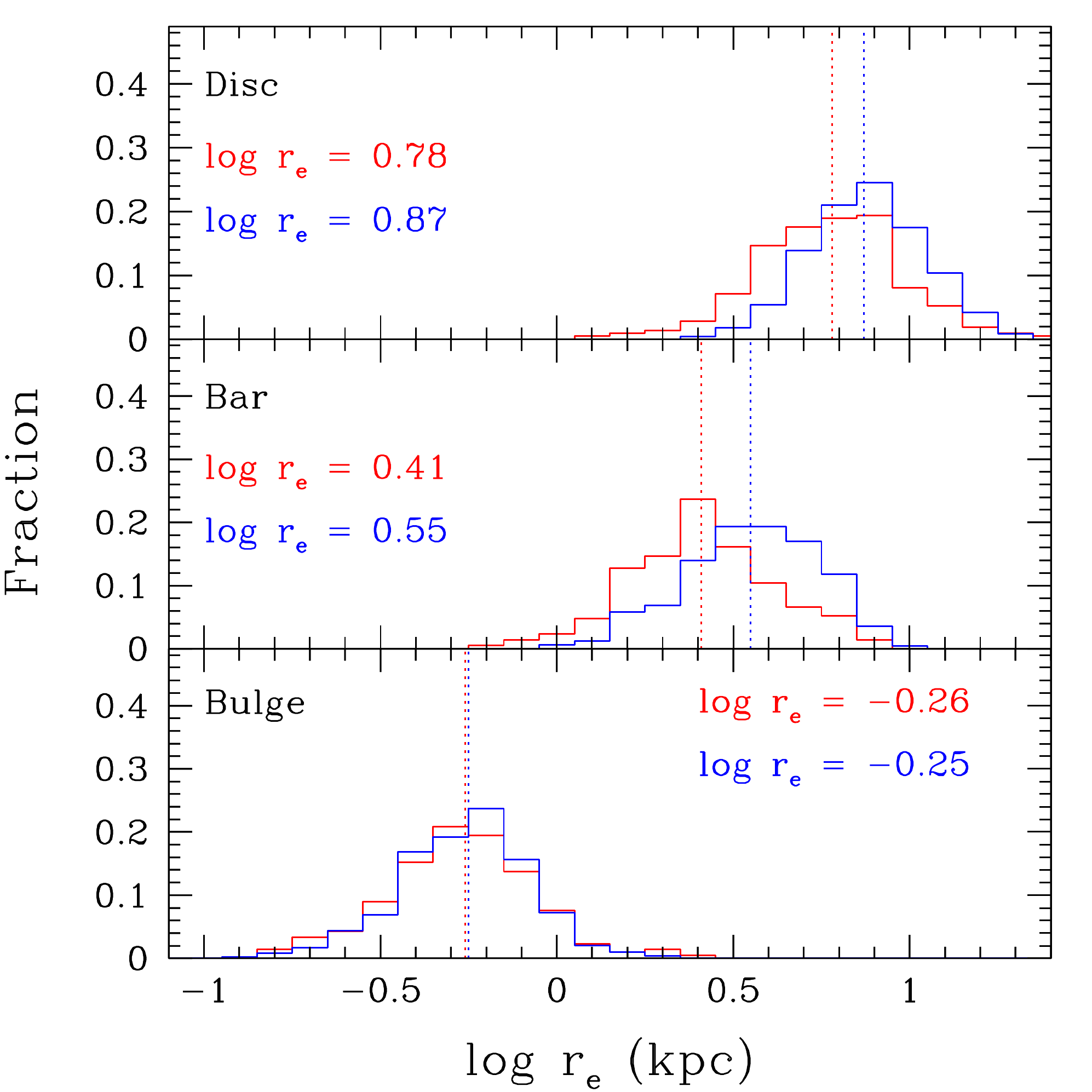}
\caption{{\bf Left:} The normalised $(g-i)$ colour distribution for bulge, bar and disc for S0 (in red) and spiral (in blue) galaxies. The vertical dashed line represents the median $(g-i)$ colours for S0 and spiral galaxies and median colours and corresponding values are shown for each component. {\bf Right:} The normalised distribution for the logarithm of bulge, bar and disc effective radii for S0 (in red) and spiral (in blue) galaxies.}
\label{fig:f1}
\end{figure*}
%%%%%%%%%%%%%%%%%%%%%%%%%%%%%%%%%%%%%%%%%%%%%%%%%%%%%%%%%%%%%%%%%%%%%%
%
%
In this paper, we use broad-band colour as a proxy to stellar population to understand the evolutionary aspect of bars by comparing them in spiral and S0 galaxies. The lowered bar frequency in S0 galaxies \citep{Butaetal2010,Nair.et.al.2010, Barway.et.al.2011} might indicate that bars in S0 galaxies are either difficult to form or bars are dissolving \citep{BournaudCombes2002,Bournaudetal2005} or bars are weaker to be detected compared to normal barred spiral galaxies. Whatever it might be, if these bars survive since their formation, one would expect them to be redder in colour in present-day galaxies. The same might hold true if a barred spiral galaxy transforms to a barred S0 as a result of a number of physical processes such as ram pressure stripping of gas and dust \citep{GunnGott1972, Larsonetal1980}; galaxy harassment \citep{Mooreetal1996} or gas starvation leading to star-formation shut down \citep{Bekkietal2002, Pengetal2015} or even mergers as shown by numerical simulations \cite[see][]{Bekkietal2011,Eliche-Moraletal2013, Tapiaetal2014, Querejetaetal2015}. So naively one would expect to find bar colours in S0s to be comparatively redder or similar to that in barred spirals. In this paper, we report, contrary to general understanding, that some bars are bluer in S0s compared to spirals and discuss its implications.

The paper is organized as follows. Section~\ref{sec:data} describes the sample that we used for this work. We present our results in section~\ref{sec:bdbarcol}, section~\ref{sec:plane} and section~\ref{sec:bluest}. Section~\ref{sec:conclu} is devoted to discussion and conclusions. Throughout this paper, we use the standard concordance cosmology with $\Omega_M= 0.3$, 
$\Omega_\Lambda= 0.7$ and $h_{100}=0.7.$

%%
%%
%%%%%%%%%%%%%%%%%%%%%%%%%%%%%%%%%%%%%%%%%%%%%%%%%%%%%%%%%%%%%%%%%%%%
\begin{figure*}
\includegraphics[scale=0.43]{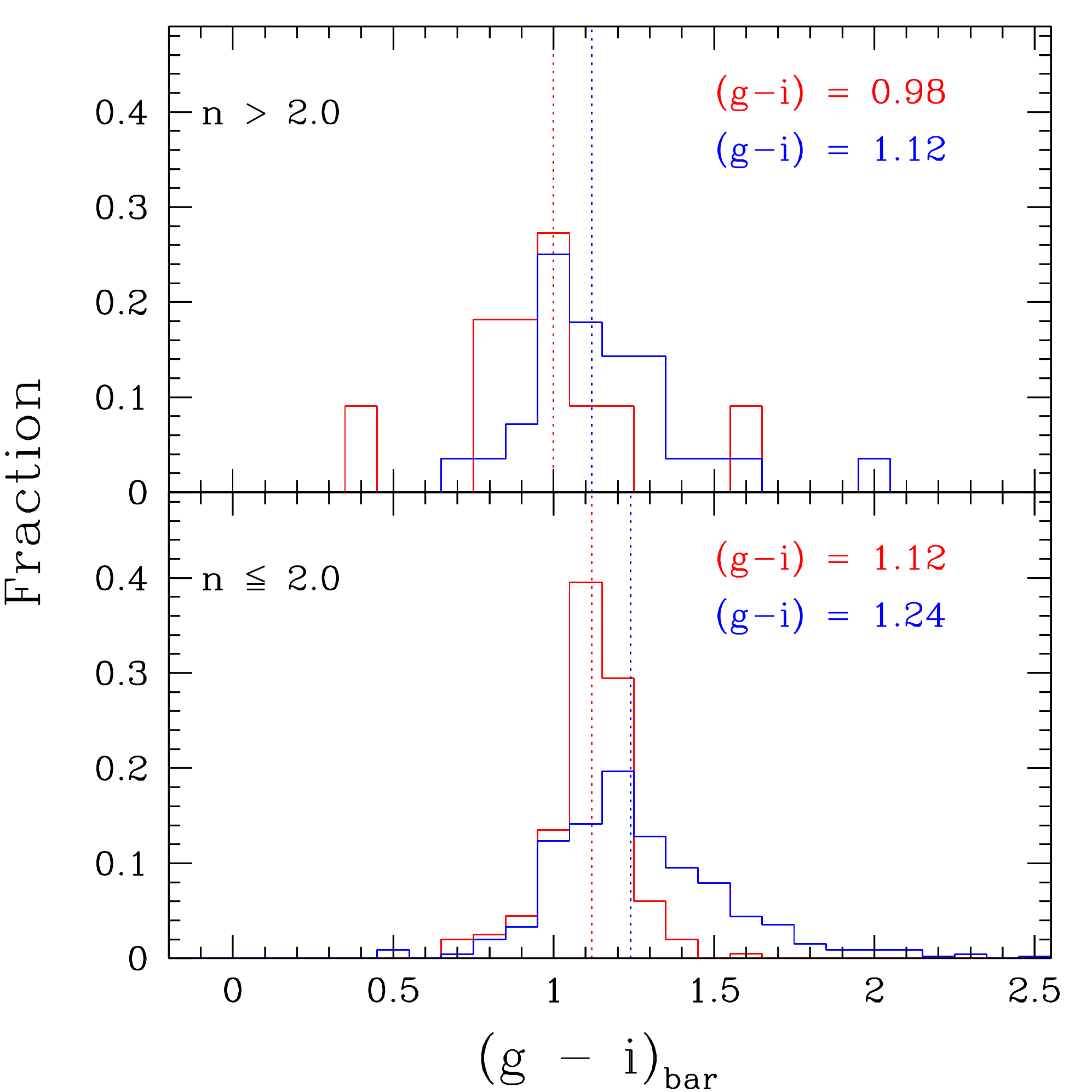}%
\includegraphics[scale=0.43]{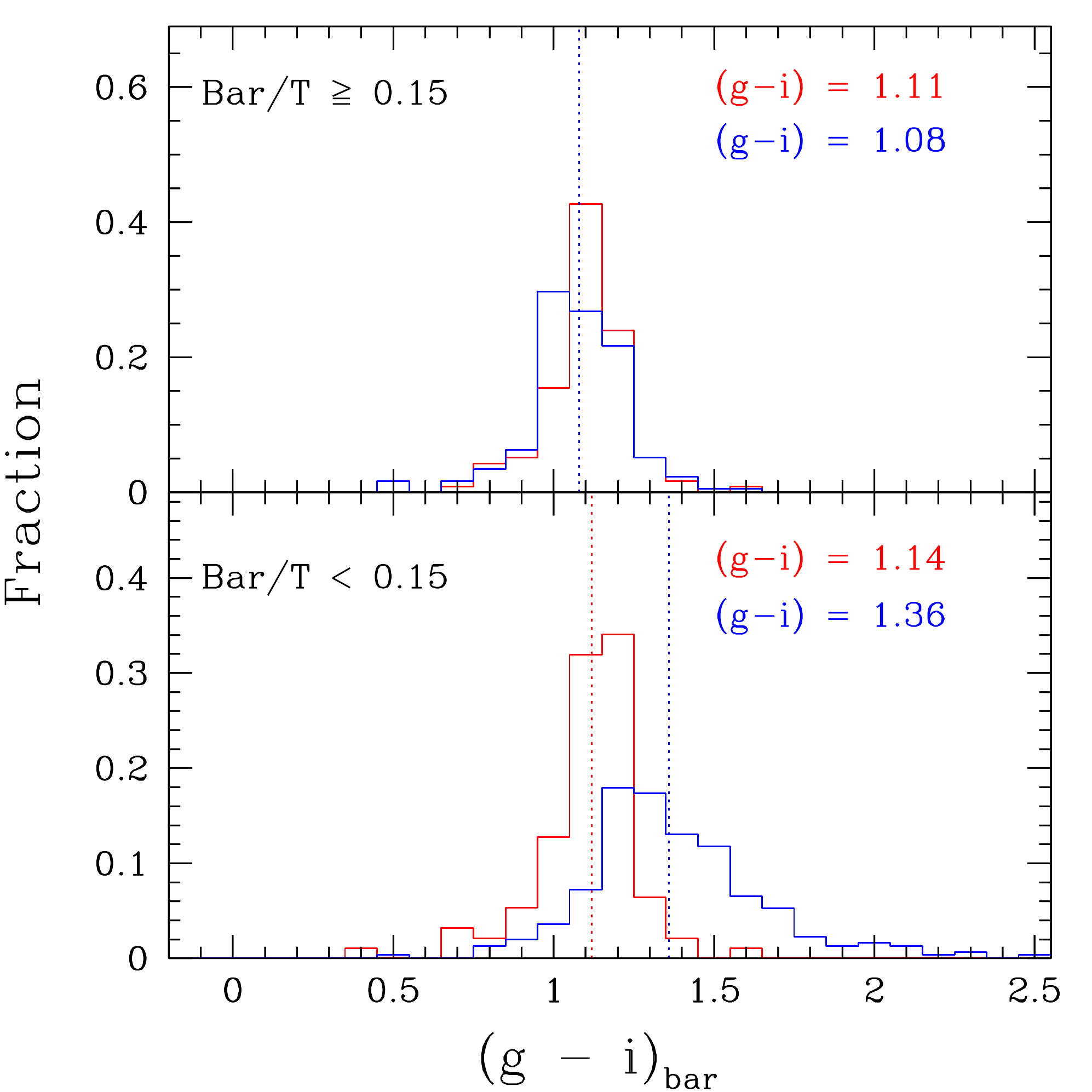}
\caption{{\bf Left:} The normalised $(g-i)$ colour distribution for the bar as a function of Sersic index 'n' for {\it classical bulges} (top panel) and {\it pseudobulges} (bottom panel) for S0 (in red) and spiral (in blue) galaxies. The vertical dashed line represents the median $(g-i)$ colours for S0 and spiral galaxies and median colours and corresponding values are shown for each component. {\bf Right:} The normalised $(g-i)$ colour distribution for the bar as a function of the bar-to-total luminosity ratio in the $i$-band. }
\label{fig:f2} 
\end{figure*}
%%%%%%%%%%%%%%%%%%%%%%%%%%%%%%%%%%%%%%%%%%%%%%%%%%%%%%%%%%%%%%%%%%%%%%%
%%
%%
%%
%%
%%%%%%%%%%%%%%%%%%%%%%%%%%%%%%%%%%%%%%%%%%%%%%%%%%%%%%%%%%%%%%%%%%%%%%
\begin{figure}
\includegraphics[scale=0.41]{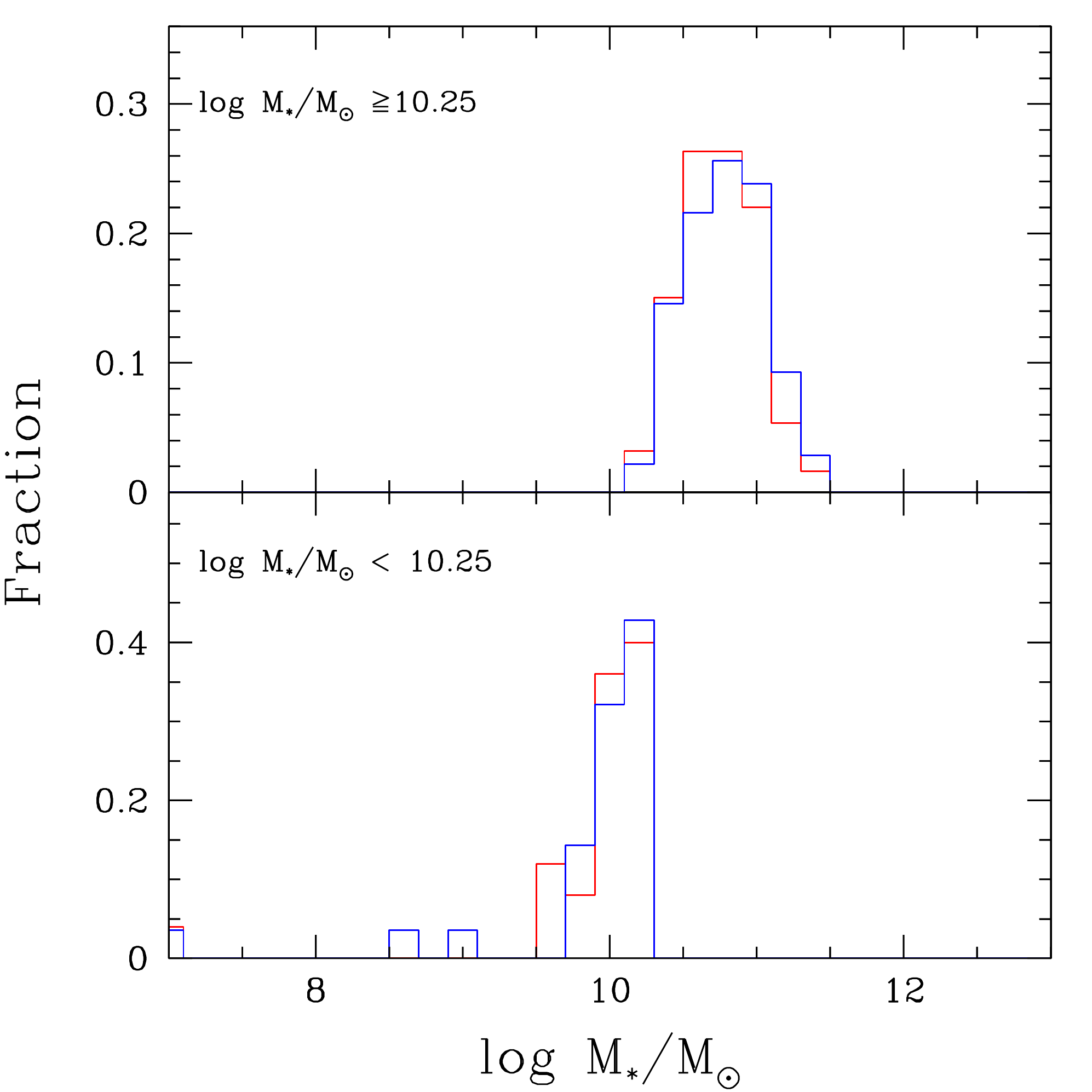}
\caption{The total stellar mass distribution for our sample of S0 (in red) and spiral (in blue) galaxies.}
\label{fig:f3} 
\end{figure}
%%%%%%%%%%%%%%%%%%%%%%%%%%%%%%%%%%%%%%%%%%%%%%%%%%%%%%%%%%%%%%%%%%%%%% 
%%
\section{Sample and analysis}
\label{sec:data}
 For this work, we use the data from \cite{Kruk.et.al.2018} (hereafter K18) which present the multi-band 2D photometric decomposition of $\sim$3500 galaxies from SDSS with strong bars. Details on data and methods can be found in K18. Here we provide a summary for the benefit of readers. K18 uses publicly available images from  SDSS  in all the five bands $u$, $g$, $r$, $i$ and $z$ for 2D photometric decomposition with the bulge, disc and bar components using GALFITM. This code was developed by the MegaMorph project \citep{Bamford2011, Haubler2013} and is a modified version of GALFIT 3.0 \citep{Peng2010}.  The advantage of using GALFITM is simultaneously fitting the wavelength-dependent model with multiple components to SDSS images in five bands. During the fitting user can vary the parameters between the bands and/or fixing some of them.  The simultaneously fitting in SDSS five bands increases the overall signal-to-noise ratio (S/N). It also uses the colour differences between the components to help with the decomposition which provides consistent colours for each component. During the fitting, K18 constrain some of the parameters, however,  the magnitude in each band was allowed to vary which help to ignore colour and, hence stellar population gradients within the independent models of each component.

{Using an iterative process, K18 fitted the barred galaxies in the sample with three components (disc+bar+bulge) for bulge dominated galaxies and two components (disc+bar) for disc dominated galaxies by adding one component at a time. The Sersic profiles were used to fit the bulge and bar components while the disc was fitted with the exponential profile. This gives magnitudes, effective radii and Sersic index for bulge and bar, magnitudes, scale lengths for the disc in each band. The magnitudes in $g$ and $i$ bands were used to obtain colours for bulge, bar and disc.

The uncertainties on the output parameters computed by GALFITM is statistical due to its assumption of the source of error is Poisson noise and are known to underestimate the real error.  It does not take into account the errors in sky estimations, the accuracy of PSF and correlated noise due to parameter degeneracy. As shown by \citet{Peng2010}, the uncertainties in estimating the sky background are one of the main source of errors. By obtaining the uncertainties in sky estimation for single Sersic fitting using GALFITM, \citet{Vikaetal2013} showed that the uncertainty in $g$ band magnitude is  $\pm$0.09 and in $i$ band magnitude is $\pm$0.11, an indicative uncertainty value for K18 measurements as they use the same software and images of the same quality. Please note the uncertainties on fitting multiple components are more complex and not trivial to estimate. For detailed discussion on this subject, we recommend reader to refer K18 and \citet{Vikaetal2013}. 

Dust reddening, extinction and K-corrections were applied to the magnitudes and colours. As mentioned in K18, the effect of internal dust reddening is not significant as galaxies in the sample is moderately face-on (i $\leq$ 60$^\circ$). We have used 2D  decomposition parameters in $g$, $r$ and $i$ bands as these are more reliable compared to decomposition in $u$ and $z$ band. These parameters for bulge dominated galaxies are used to explore the colour distribution of bars, bulges and discs of sample galaxies as a function of galaxy morphology. The differences in component colours and trends with stellar mass and environment will be useful to investigate the bar properties, particularly in S0 galaxies.

Next, we search for reliable visual morphological classification for galaxies in K18  to have a sample of S0 and spiral galaxies, respectively. For this purpose, we cross-match K18  sample of barred galaxies with \cite{Nair.et.al.2010} catalogue which provides a careful morphological classification using SDSS images in all the five bands $u$, $g$,$r$, $i$, and $z$. This catalogue can differentiate between S0 galaxies into S0-, S0, and S0/a and spirals galaxies into Sa, Sab, Sb, Sbc, Sc, Scd, Sdm, Sm, and Im.  The galaxies in \cite{Nair.et.al.2010} catalogue have been classified twice and have estimated a mean deviation of less than 0.5 T-types. In this work, we group together all types of S0 galaxies (S0-, S0, and S0/a) and refer them as S0s. We do the same to have a spiral galaxy sample.

This cross-match gives 692 galaxies with bulge, disc and bar parameters.  We find that out of  694, the number of S0 and spiral galaxies are 211 and 481, respectively.  In this spiral galaxy sample, there are 428 early-type spiral galaxies (ranging from Sa-Sbc types) and 53 late-type spiral galaxies (ranging from Sc-Sd types) which make it dominated by early-type spiral galaxies.

%%
%
%%
%%%%%%%%%%%%%%%%%%%%%%%%%%%%%%%%%%%%%%%%%%%%%%%%%%%%%%%%%%%%%%%%%%%%%%
\begin{figure*}
\includegraphics[scale=0.29]{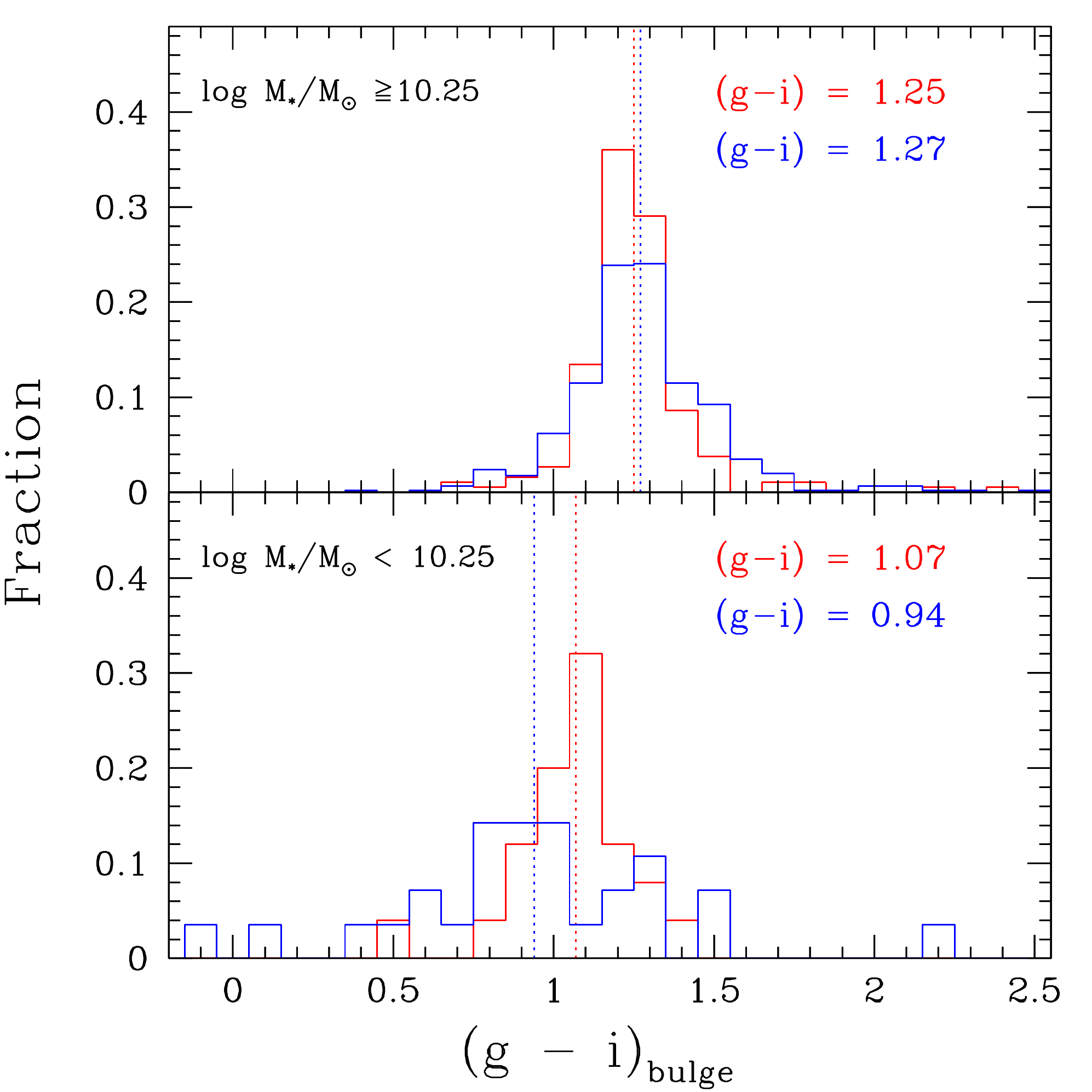}%
\includegraphics[scale=0.29]{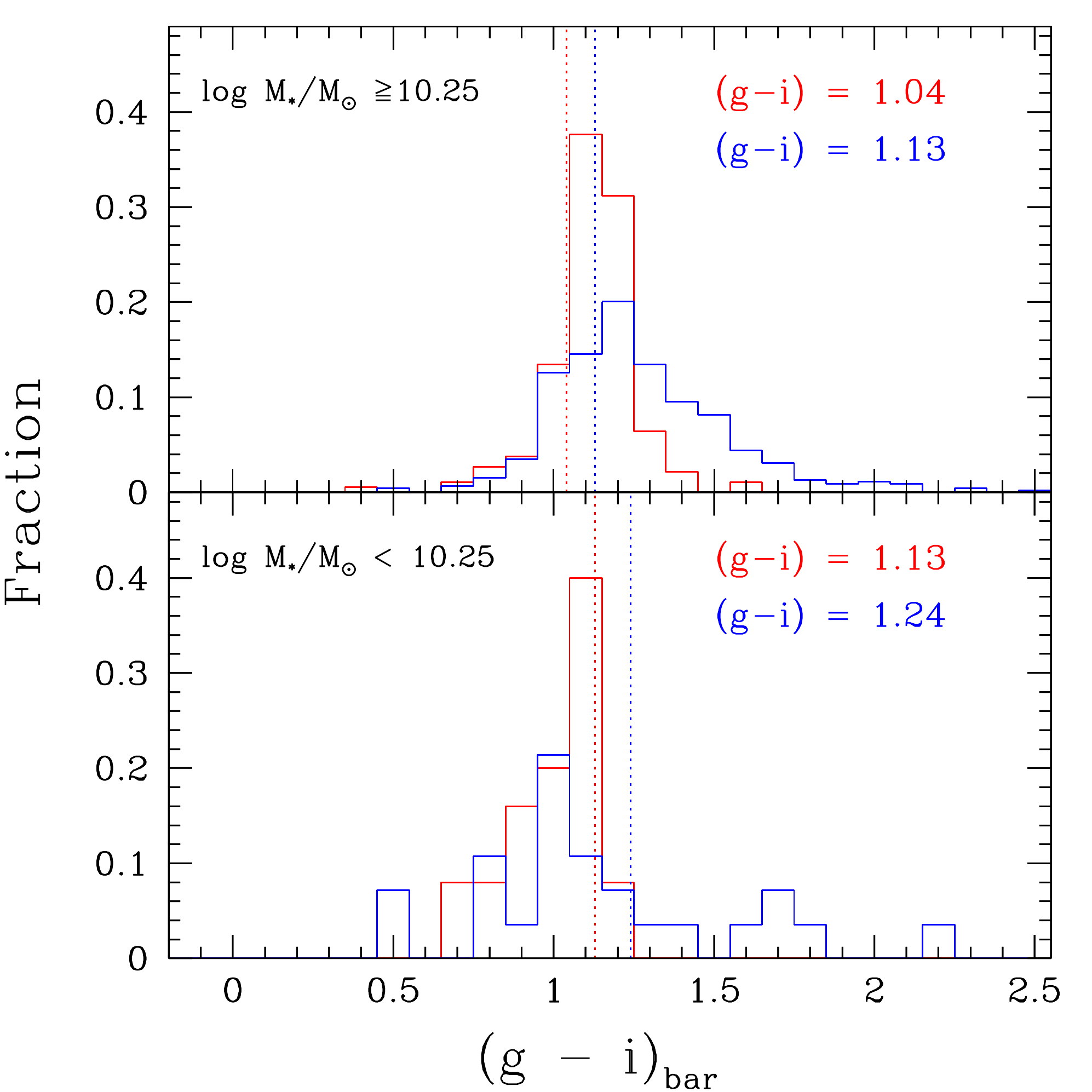}%
\includegraphics[scale=0.29]{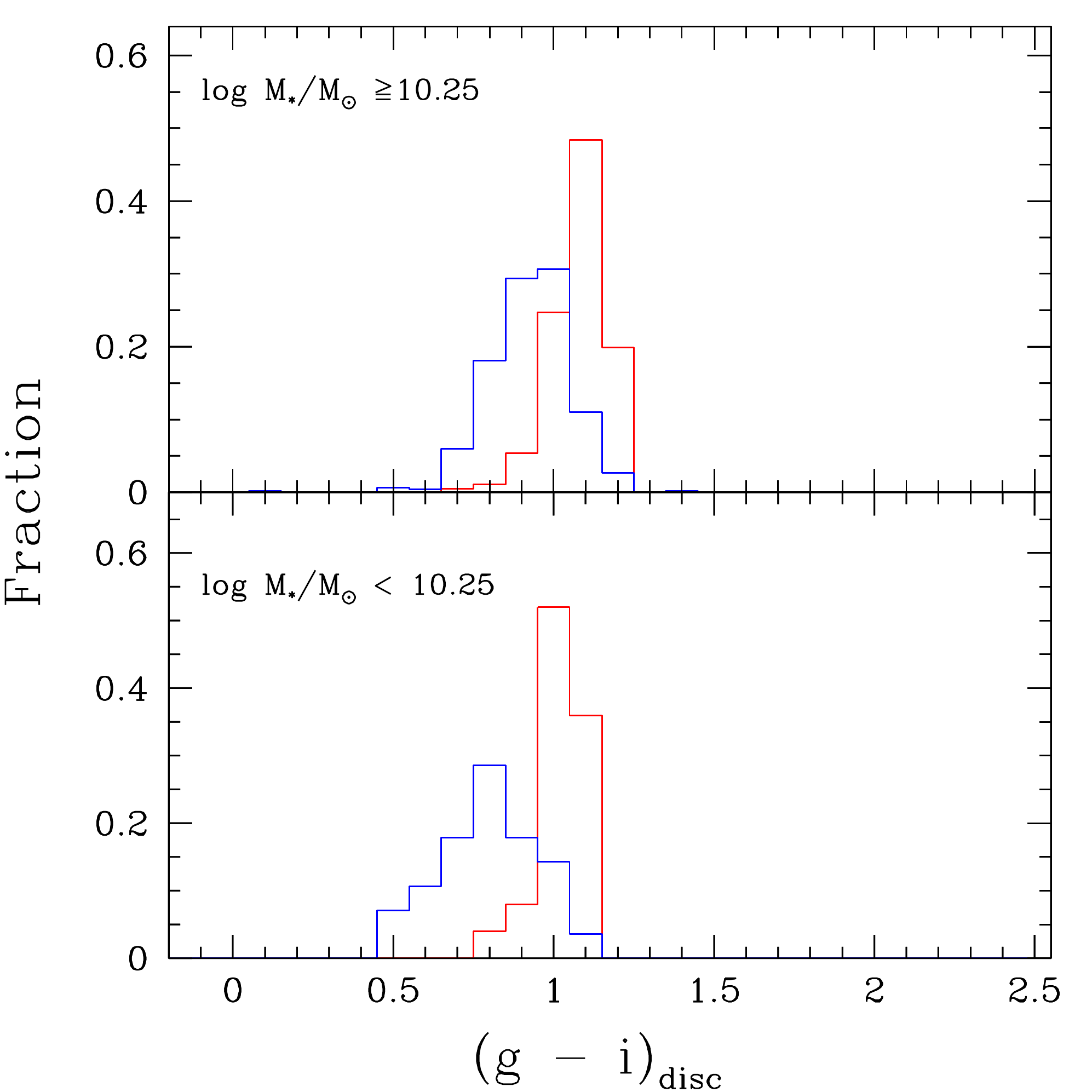}
\caption{The normalised $(g-i)$ colour distribution for bulge (left), for bar (middle) and for disc (right) plotted as a function of total galaxy mass for S0 (in red) and for spiral (in blue) galaxies. We have used the 10$^{10.25}$ M$_\odot$ as a mass division similar to K18 (see text for details).}
\label{fig:f4} 
\end{figure*}
%%%%%%%%%%%%%%%%%%%%%%%%%%%%%%%%%%%%%%%%%%%%%%%%%%%%%%%%%%%%%%%%%%%%%% 
%
%

\section{Bulge, disc and bar colour distribution}
\label{sec:bdbarcol}
 \cite{Kruk.et.al.2018} have shown using $(g-i)$ colour distribution for bulge, disc and bar that the discs are bluer than the bars and bulges while there is little difference between the bar and bulge $(g-i)$ colour.  The median difference between bulges and discs for $(g-i)$ is 0.33 and between the bars and discs is 0.20. The authors used $(g-i)$ colour as these two bands are sufficiently separated in wavelengths and less affected by dust extinction. 

We show the distribution of optical colours ($g-i$) for the bulge, disc and bar components in Fig.~\ref{fig:f1} as a function of morphology for S0 and spiral galaxies. While the bulge colour shows no difference, discs in S0 galaxies are redder compared to spiral galaxies with median $(g-i)$ colour difference of 0.15 magnitudes. The bluer discs in spiral galaxies are as per norm \citep{Bakosetal2008,MacArthuretal2009}; however, the distribution of the bar colours in S0s and spiral galaxies are unexpected \citep{Tullyetal1982, Buta.et.al.2010, Lansbury.et.al.2014}. Bars in the current sample of S0 galaxies are bluer compared to spiral galaxies with a median $(g-i)$ colour difference of 0.11 magnitude. However, is the  difference between bar colours for S0 and spiral galaxies significant? To address this, we perform a statistical test. A commonly used non-parametric test is the Kolmogorov–Smirnov test (hereafter K–S test), which measures the maximum distance ($d$) between the two cumulative distributions. Under the null hypothesis that the two samples tested are drawn from the same population, a larger maximum distance is less likely to occur.  The K–S test reveals that the $(g-i)$ colour distribution for S0 and spiral is drawn from the same distribution is rejected ($d$ = 0.40) with a significance level ($P$) of 10$^{-5}$ or better. 

If we consider a scenario in which a barred S0 galaxy is transformed from a barred spiral galaxy by the mechanisms involving no past major interaction that does not destroy the bar, the stellar content of S0 bars should be dominated by old stars. As a result, one would expect the bar colour to be redder in S0 galaxies compare to those in spirals. However, Fig.~\ref{fig:f1} reveals an opposite picture. When we compare the $(g-i)$ colour distribution of the bulge, disc and bar components for S0 as an individual morphology class, we see that the discs and bars are bluer compared to bulges; with discs being marginally bluer than bars. On the other hand, for spiral galaxies, the $(g-i)$ colour distributions for the bulges and bars are similar and discs are bluer as one would expect normally\citep{Bakosetal2008,MacArthuretal2009}. In other words, barred spiral galaxies have an older stellar population in the bar and bulge \citep{Sanchez-Blazquezetal2011}. This brings out a puzzling issue in which bars are bluer and probably star-forming in S0 galaxies - something that is contrary to our overall understanding of galaxy dynamics and evolution \citep{Sellwood2014,Conselice2014}.

The right panel of Fig.~\ref{fig:f1} shows the distribution of the effective radii of bulge, bar and disc of S0 (in red) and spiral (in blue) galaxies. The bulge effective radii for S0s and spiral galaxies span a similar range while the bar, as well as disc effective radii, are smaller in S0s compared to spirals. This suggests that our S0 galaxies have smaller bars compare to their spiral counterpart. A two-sample K-S analysis reveals that the distribution of bar and disc effective radii for S0 and spiral is drawn from the same distribution is rejected ($d$ = 0.36) with a significance level ($P$) of 10$^{-5}$ or better. Note that in the current sample of S0 and spiral galaxies, bulges are, in general, smaller and compact compared to bars whereas discs are bigger.

% done here.
%%%%%%%%%%%%%%%%%%%%%%%%%%%%%%%%%%%%%%%%%%%%%%%%%%%%%%%%%%%%%%%%%%%%%%
\begin{figure*}
\includegraphics[scale=0.95]{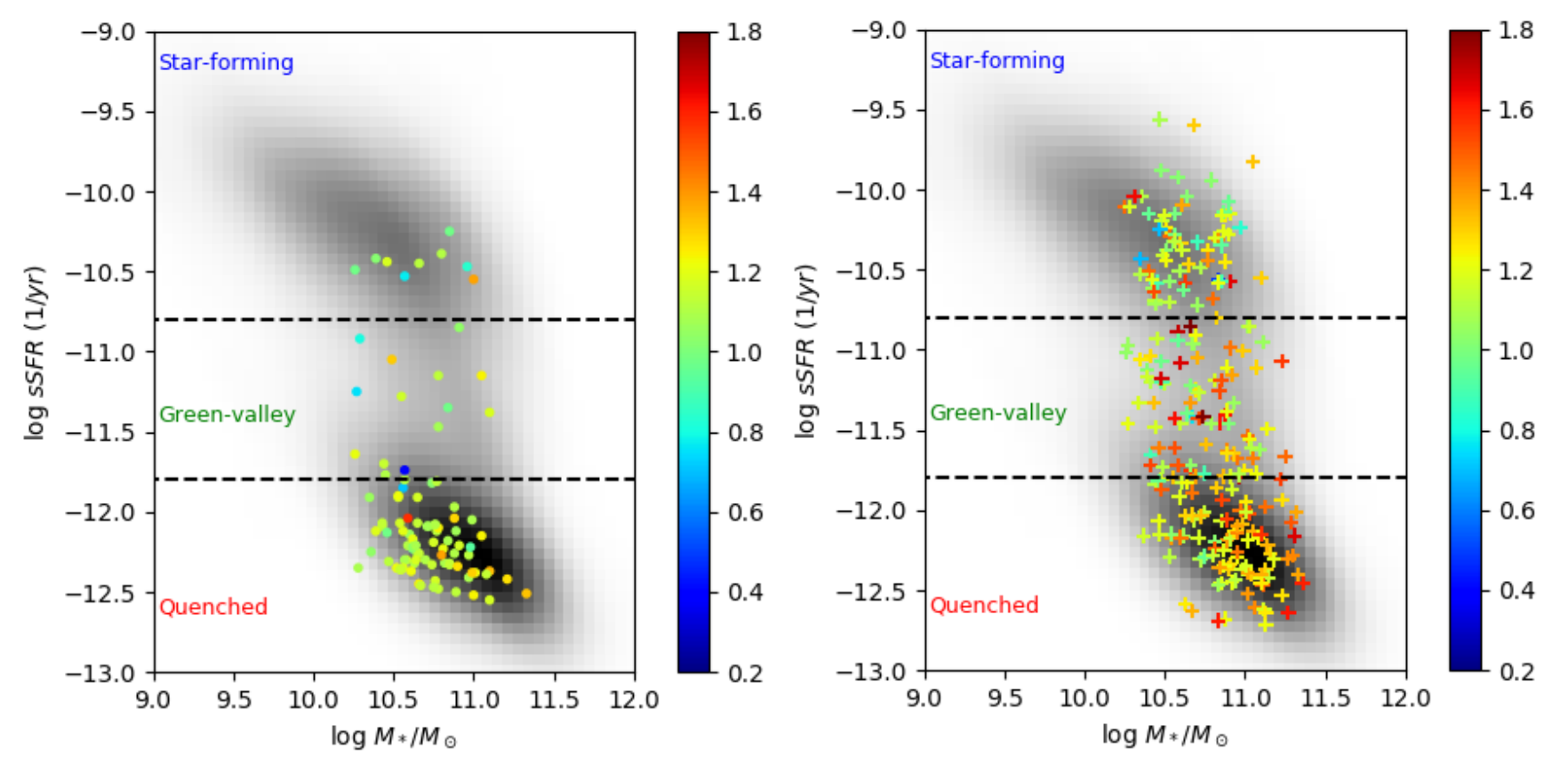}
\caption{Dependence of specific star formation rate on stellar mass for S0 (left panel) and for spiral (right panel) galaxies as a function of bar colour for massive galaxies ($M_* > 10^{10.25} \ M_\odot$). The region between the horizontal blue dashed lines defines the green valley from \citet{Salim.2014}. Galaxies above the upper blue line ($log \ sSFR = -10.8$) are termed star forming, and those lying under the lower green line ($log sSFR = -11.8$) are termed quenched galaxies. The grey background density of galaxies is from  \citet{Bait.et.al.2017} which is available publicly for all morphological classes. The S0 galaxy SDSS J123553.51+054723 having bluest bar (shown as blue dot) is at boundary of green valley towards quenched side (left panel).}
\label{fig:f5} 
\end{figure*}
%%%%%%%%%%%%%%%%%%%%%%%%%%%%%%%%%%%%%%%%%%%%%%%%%%%%%%%%%%%%%%%%%%%%%%
%
%
\subsection{Role of bulge and disc on bar properties}
\label{sec:bdrole}
It will be interesting to consider the role of a bulge and disc to understand comparatively bluer colours of bars in S0 galaxies. Lets first explore the role of discs in bar formation. A number of simulations have shown that the massive, cool, self-gravitating discs are prone to bar instability which leads to the formation of strong bars \citep{CombesSanders1981,SellwoodWilkinson1993,Athanassoula2002,Dubinskietal2009,SahaElmegreen2016} whereas lower mass, comparatively hotter galaxies do not form strong bars, and the bar develops over a longer period of time \citep{Sahaetal2010, sheth12, Saha2014}. On the observational side, it has been shown that the incidence of a bar in S0 galaxies does not depend on the stellar mass of the host discs \citep{Barway.et.al.2016}.  The $(g-i)$ colour of the disc for S0 is redder compare to spiral galaxies and also the disc effective radii are smaller in S0s compared to spiral galaxies. This shows the possible evidence of secular evolution via bars, quenching of star formation in the discs and possible route to the formation of pseudobulges \citep{KK2004}.

As shown in K18, the dominant fraction of bulges in these galaxies are of pseudobulge type as found in other studies \citep{Jogeeetal2005,Kormendy.2013}. There are a very few elliptical-like {\it classical bulges} in this sample. K18 has used Sersic index 'n' for identifying bulges in which bulges with $n \le 2$ are classified as {\it pseudobulges} and with $n > 2$ are classified as  {\it classical bulges} \citep{FisherDrory2008}. This identification scheme is still a matter for debate in the literature \citep{Gadotti2009,Vaghmare2013}; although, for a large sample, researchers generally use it. In this work, we make use of this division for comparing our results with K18 and with different studies in the literature. This division gives us 200 (95~\%) {\it pseudobulges} in S0s and 453 (94~\%) pseudobulges in Spiral galaxies. Rest of the bulges, 11 (5~\%) in S0 and 28 (4~\%) in spiral galaxies are {\it classical bulges}. Given that our sample is dominated by {\it pseudobulges}, see also \cite{Weinzirletal2009}; it will be interesting to explore the $(g-i)$ colour distribution as a function of bulge type. The right panel of Fig.~\ref{fig:f2} shows the normalised $(g-i)$ colour distribution for Sersic index 'n'; top panel shows the distribution for {\it classical bulges} while the bottom panel shows the distribution for {\it pseudobulges}. It is clear that for both types of bulges, median $(g-i)$ colour for the bar is comparatively bluer in S0s than in spirals. Amongst the S0 galaxies, the median $(g-i)$ colour of the bar in {\it classical bulges} is bluer than for the bar in {\it pseudobulges}. It is worth to note that the bluest and reddest bar colour for our sample of S0 galaxies host {\it a classical bulge}.

As mentioned earlier, the bulges of barred galaxies for both S0 and spiral galaxies are predominantly of pseudo-bulge type and are a likely result of the secular evolution processes. The {\it classical bulges}, which are believed to have formed by major and minor merger events \citep{Aguerrietal2001,Hopkinsetal2010} are uncommon in our sample of S0 and spiral galaxies and have bluer bars particularly in S0 galaxies. It remains intriguing to explore what make bars bluer in S0 galaxies hosting {\it classical bulges}. 

Since this sample consists of strongly barred galaxies, it is imperative to examine the bar prominence to understand the bar colour difference we see in S0s and spirals. We use bar-to-total luminosity ratio (Bar/T) to indicate the bar strength or the mass of a bar \citep[see also][]{Weinzirletal2009}. \cite{Kruk.et.al.2018} measured the Bar/T ratio for their sample of galaxies in all five bands of SDSS with a median value of $\sim 0.14$ in $i$-band for bulge dominated galaxies. The right panel of Fig.~\ref{fig:f2} shows the $(g-i)$ colour distribution for the bar as a function of Bar/T ratio. We use $Bar/T = 0.15$ to divide our sample into strong bars and less-strong bars for S0 and spiral galaxies. For $Bar/T > 0.15$ i.e., massive bars, there is no difference in bar colour for S0 and spiral galaxies suggesting of having a similar stellar population for the stronger bar in both morphological types. The bar colour difference is more prominent for $Bar/T \le 0.15$ (according to K-S test; d=0.60 with a significance level (P) of $10^{-5}$ or better) implying that the less-strong and probably smaller-sized bars are bluer in S0s compared to spiral galaxies. Now, it remains to be investigated what makes smaller bars appear bluer in S0s.

\subsection{Role of stellar mass}
\label{sec:mass}
Dependence of galaxy properties such as size, colour, star formation rate on stellar mass is well known \citep{Conselice2006}. K18 has explored this mass dependence for their sample which contain the galaxies of stellar masses from 10$^8$~M$_\odot$ to 10$^{11.5}$ M$_\odot$. They have shown that the colours of the discs and bars for low-mass galaxies are bluer compared to massive galaxies when using 10$^{10.25}$ M$_\odot$ as mass division.

For massive galaxies ($M_{*} > 10^{10.25} \ M_\odot$), the bulge is redder and bulge and bar colours are similar suggesting old stellar populations in both components as can be seen in simulated galaxies from cosmological hydrodynamical simulations \citep{Scannapiecoetal2010}.  We are using a sub-sample from K18 obtained by morphological classification in S0 and spiral galaxies and investigate the stellar mass dependence in Fig.~\ref{fig:f3} which shows the stellar mass distribution for S0 and spiral galaxies. For high and low mass, both morphological classes have a similar span of stellar masses despite having different sample size for S0 and spiral galaxies. Using the same mass division given in K18, we plot the normalised distribution of $(g-i)$ colour for the bulge (left panel), disc (right panel) and bar (middle panel) components in Fig.~\ref{fig:f4}. Interestingly, the bars in massive S0 galaxies are bluer compared to massive spiral galaxies (according to K-S test; d=0.39 with a significance level (P) of $10^{-5}$ or better). The scenario is the same for the low mass S0 galaxies in that bars are bluer compared to their spiral counterpart. Interestingly, if we consider only the S0 galaxies in our sample,  massive S0 galaxies have the bluer bar colour compare to less massive S0 galaxies (according to K-S test; d=0.36 with a significance level (P) of $10^{-5}$ or better). Also, most of the {\it classical bulges} hosting S0 galaxies are massive.  This indicates that the stellar population of bars in massive S0 galaxies is younger than the spiral galaxies of similar mass range and less massive S0 galaxies. 

The $(g-i)$ colour using this mass division for bulge (left panel) and the disc (right panel) is shown in Fig.~\ref{fig:f4}. The disc colours show the expected behaviour i.e. spiral discs are bluer than S0 discs, however, this difference is more pronounced for low mass galaxies. We do not see a difference for the bulge $(g-i)$ colour for the massive S0 and spiral galaxies. For both morphological types, the bulge colour is redder than the bar colour. This probably implies that the bulge does not play a role in turning the bar appear bluer. Though the bulge $(g-i)$ colour for low mass galaxies is bluer compared to bar colour and this difference is more prominent for low mass spiral galaxies. 

It is worth to mention here that when the mass division given in K18 is used, the sample size of our galaxies towards the low mass end is considerably less ($\sim$8\%). The morphological type for most of these low mass spiral galaxies is late-type (Sc and later). This explains the difference between $(g-i)$ colour for the bulge, bar and disc of low mass galaxies for spiral and S0s. The discs of low mass late-type spiral galaxies are more gas-rich and hence star-forming \citep{Kennicutt1983,Zhang.et.al.2018}.  The bulges of low mass late-type spiral galaxies are bluer than S0 again showing typical behaviour. However, the $(g-i)$ colour of the bar is redder for low mass late-type spiral galaxies compared to S0s.  This kind of behaviour is only possible when a galaxy is formed through a secular process in which the discs were formed first and bulges emerged from the disc through the gas transported by the bar and likely to be significant in late-type spiral galaxies \citep{KK2004}. 

\section{Barred galaxies on \texorpdfstring{sSFR-M$_{*}$} \ plane}
\label{sec:plane}
It will be interesting to investigate the position of our sample of barred galaxies on the sSFR-M$_{*}$ plane for massive galaxies ($M_* > 10^{10.25} \ M_\odot$) as these galaxies are dominant in our sample. \citet{Bait.et.al.2017} has found using a sample of massive galaxies  ($M_* > 10^{10} \ M_\odot$) that the green valley is mostly populated by early-type spiral and S0 galaxies. These authors suggest that morphology plays an important role in determining the star-forming state of massive galaxies in the local universe; although, they do not consider the role of features like bars and rings. Recently \citet{Kelvin.et.al.2018} find a significant fraction of rings and lenses in disc-type galaxies in the green valley with a presence of the bar common in most of these galaxies. K18 has shown that the barred disc galaxies are more common in green valley and quenched sequence compared to unbarred galaxies at a given mass. We explore the same by dividing our sample of massive barred disc galaxies into S0s and spiral galaxies on the sSFR-M$_*$ plane as a function of bar colour with the galaxies common with publicly available data from \citet{Bait.et.al.2017}. These authors used multi-wavelength photometry, from UV-optical-near-mid IR, for $\sim$6000 galaxies in the Local Universe to model the Spectral Energy Distribution (SED) to estimate the stellar mass and SFRs. Only optical imaging data from the SDSS  in $u, g, r, i$, and $z$ bands for all our sample galaxies can be used for SED fitting; however, the UV and Mid-IR data provides a better constraint on recent star formation and dust estimation during SED fitting.  \citet{Bait.et.al.2017} have used aperture matched total magnitudes in UV-optical-IR bands to model the SED, using the publicly available Multi-wavelength Analysis of Galaxy Physical Properties (MAGPHYS) code \citep{daCunhaetal2008} to estimate the stellar mass ($M_*$)and SFRs with typical uncertainty in log $M_*$ is about 0.048 dex and in log sSFR is about 0.125 dex. The cross-matched of our sample used in this work with \citet{Bait.et.al.2017} catalogue gives 101 S0 and 243 spiral galaxies. 
%
%
%%%%%%%%%%%%%%%%%%%%%%%%%%%%%%%%%%%%%%%%%%%%%%%%%%%%%%%%%%%%%%%%%%%%%%
\begin{figure}
\begin{center}
\includegraphics[scale=0.41]{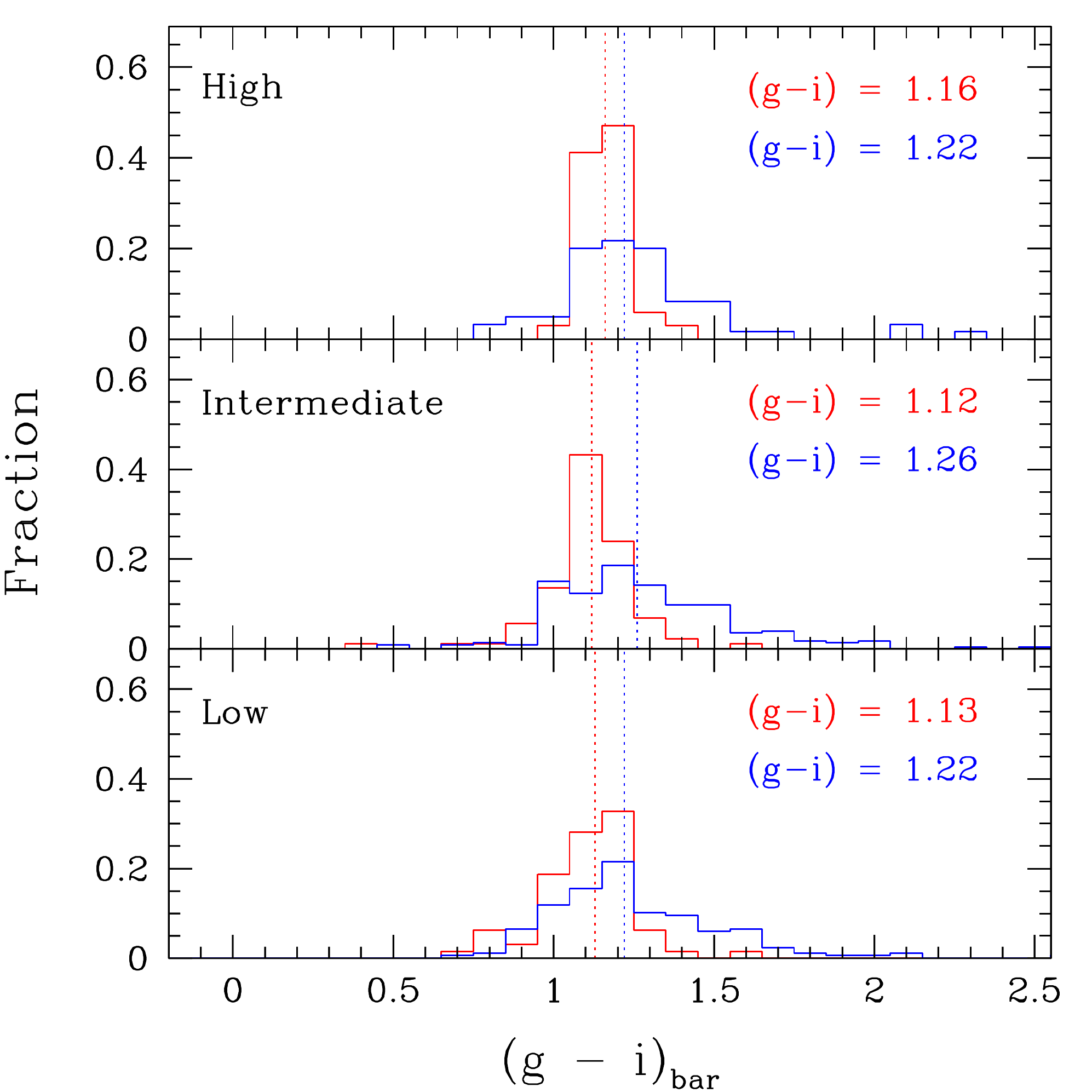}
\caption{The normalised distribution of (g-i) bar colour for S0 (in red) and for spiral (in blue) galaxies as a function of local environmental density ($\log \ \Sigma \ (\text{Mpc}^{-2})$)(see text for details).}
\label{fig:f6}
\end{center}
\end{figure}
%%%%%%%%%%%%%%%%%%%%%%%%%%%%%%%%%%%%%%%%%%%%%%%%%%%%%%%%%%%%%%%%%%%%%%
\begin{figure*}%[htb!]
\begin{center}
\includegraphics[scale=0.65]{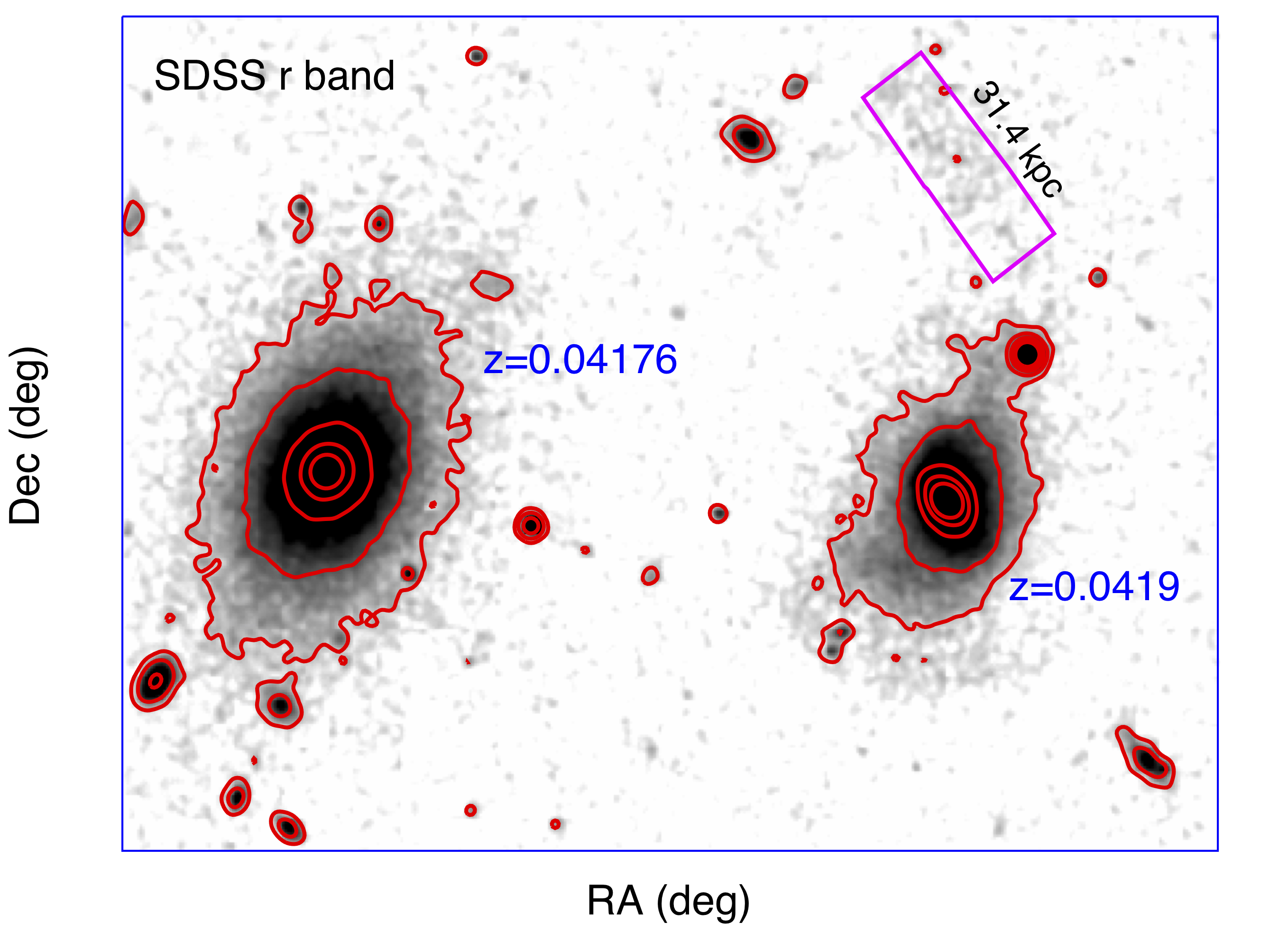} 
\caption{$r$ band images of S0 galaxy SDSS J123553.51+054723 having bluest bar. The outermost contour is at $1\sigma = 0.0253$, subsequent inward contours are drawn at $3\sigma$, $10\sigma$,$20\sigma$, $40\sigma$. The S0 disc (SDSS J123553.51+054723, galaxy on the right) showing disturbed morphology and signs of interaction possibly with the companion (left) at a projected separation of 700 kpc. The total counts in the magenta box of length 31.4 kpc is 38.4 which is about 6 times more than the background.}  
\label{fig:f7}
\end{center}
\end{figure*}
%%%%%%%%%%%%%%%%%%%%%%%%%%%%%%%%%%%%%%%%%%%%%%%%%%%%%%%%%%%%%%%%%%%%%%
%%
%%
 In Fig.~\ref{fig:f5} we show the dependence of specific star formation rate on stellar mass for S0s (left panel) and for spiral (right panel) galaxies as a function of bar colour for massive galaxies ($M_* > 10^{10.25} \ M_\odot$). The region between the horizontal dashed lines defines the green valley from \citet{Salim.2014}. Galaxies above the upper black line ($log \ sSFR = -10.8$) are termed star-forming, and those lying under the lower dashed line ($log sSFR = -11.8$) are termed quenched galaxies. The grey background density of galaxies is from  \citet{Bait.et.al.2017} which is available publicly for all morphological classes. 

Note that in the Fig.~\ref{fig:f5}, the number of spiral galaxies in the ‘star-forming main-sequence’ is much higher than the S0 galaxies as spiral galaxies are almost double in the number than S0 galaxies. Also as shown in Table 3 of  \citet{Bait.et.al.2017}, the fraction of spiral galaxies in 'star-forming main-sequence’ is large compared to S0 galaxies. The S0 galaxy SDSS J123553.51+054723 having the bluest bar (shown as a blue dot) is at the boundary of the green valley towards the quenched side (left panel).

It is clear from Fig.~\ref{fig:f5},  spiral galaxies in our sample populate all the region of the plane with the presence of reddest barred spiral galaxies are in the green valley. However, S0 galaxies in which most of the bars are bluer compared to spiral galaxies in this sample are in the quenched region and green valley. The bluest barred S0 galaxy  SDSS J123553.51+054723 (see Section~\ref{sec:bluest} for more details.) is in the green valley along with few more S0 galaxies with a blue bar. We have also noticed that the spiral galaxies with the blue bar are mostly in Star-forming region. If we consider a scenario where S0 galaxies are formed via stripping gas from the disc of spiral galaxies \citep{GunnGott1972, Mooreetal1996} given that the bulges of both types of galaxies are old, a burst of star formation in a bar can move S0 galaxies from quenched region to green valley. It will be interesting to investigate further what could have caused such a burst of star formation in the bar given that the SSP models with an instantaneous burst of star formation 8 Gyr ago could produce  (g - i) colour of 1.13 \citep{Fernandez.et.al.2014} which is close to the median colour of the bar in our S0 galaxy sample. The results presented in the section provide us a clue on bar evolution in massive S0 galaxies where most of the bars are shorter and comparatively bluer than spiral galaxies. 

\section{Bluest S0 bar}
\label{sec:bluest}
In this study, we found that a galaxy SDSS J123553.51+054723.5 hosting one of the bluest bar in our S0 galaxy sample with bar $(g-i)$ colour of 0.40, while bulge which is a {\it classical bulge} has $(g-i)$ colour of 1.05 and disc $(g-i)$ colour is 0.95. This galaxy is at redshift 0.042 and has an extended disturbed disc along with the tail (see Fig.~\ref{fig:f7}), a sign of past interaction, possibly with a nearby elliptical galaxy SDSS J123553.79+054539.8 at the same redshift. The HI gas was detected for SDSS J123553.51+054723.5 in the extended GALEX Arecibo SDSS Survey (xGASS) \citep{Catinella2010} which gives the integrated HI mass $ \log (M_{HI}/ M_{\odot}) =8.86$. Considering the stellar mass of this galaxy, this implies $M_{HI}/M_{*} \simeq 0.02$. Although compared to normal spiral galaxies the gas to stellar mass ratio is significantly lower, the galaxy is not literally devoid of gas as in many massive S0 galaxies. Since the galaxy seems to be interacting with the neighbour or it could be even in flybys, it might be responsible in triggering star-formation in the central region of this S0 galaxy and might even have led to the formation of the bar \citep{Lokas2018}. Since the galaxy's stellar mass is $4 \times 10^{10}$~M$_{\odot}$, it remains unclear whether the central star-formation was due to the accretion of fresh gas \citep{DekelBirnboim2006}. Other possibility might be that the interaction or the flyby could excite a bar in the central region as well as ignite star-formation from the residual gas. 

\section{Discussion and Conclusions}
\label{sec:conclu}
While bar formation is discussed in detail in the literature \citep{SahaNaab2013}, the fate of bars is still unclear in terms of observational evidences. Barred S0 galaxies can be an ideal case to study the fate of bars in the scenario where S0s are thought to have transformed from spiral galaxies via a number of processes such as ram pressure stripping. Some numerical simulations have shown that a bar may be destroyed or weakened by strong gas inflow \citep{Bournaudetal2005} while some others have shown that it can survive for a long time \citep{Athanassoula2005}.

Most of S0 galaxies in our sample have small or weak bars. One possibility is that a weak or small bar may dissolve into a lens feature as proposed \citep{Kormendy1979, Kruk.et.al.2018} which interestingly also found that the $(g-i)$ colour for the lenses is bluer compared to bars. Moreover, the bar and lens fraction in S0 galaxies are similar as given in \citet{Nair.et.al.2010} indicate the possibility of bars can dissolve into lenses for S0 galaxies. 

Using H$_\alpha$ data, star formation in the bar region for spiral galaxies have been reported in the literature \citep{Martin.et.al.1997, Sheth.et.al.2002}. Simulations have been shown that at the edge of the bar region conditions might be favourable for the formation of molecular complexes and thus mini-starbursts \citep{Renaud.et.al.2015}, however, also for spiral galaxies. There is no such investigation have been done to study star formation in bar region of S0 galaxies.  This might be due to the fact that most of the S0 galaxies host small or weak bar. However, star formation can be triggered by a weaker shock due to some interaction when a galaxy is a part of a group or cluster. \citet{Barway.et.al.2011} have shown that a significant number of barred S0 galaxies are part of a group/cluster environment. For this sample of barred S0 galaxies, we use local environmental density ($\log \ \Sigma \ (\text{Mpc}^{-2})$) as given in  \citet{Nair.et.al.2010}. We split our sample of S0 galaxies into  low density ($\log \Sigma \ (\text{Mpc}^{-2}) < -0.5$),  intermediate-density (-0.5 < $\log \Sigma \ (\text{Mpc}^{-2}) < 0.5$), and high densities ($\log \Sigma \ (\text{Mpc}^{-2}) > 0.5$) similar to \citet{Bait.et.al.2017}. A distribution of (g-i) bar colour as a function of local environmental density ($\log \ \Sigma \ (\text{Mpc}^{-2})$) is shown in Fig.~\ref{fig:f6} indicating that a significant fraction of barred S0 galaxies is in the intermediate-density environment. The median $(g-i)$ colour difference 0.14 magnitudes for bars in S0 galaxies are bluer compared to spiral galaxies in the intermediate-density environment (according to K-S test; d=0.45 with a significance level (P) of $10^{-5}$ or better). Though the bars in S0 galaxies are bluer compared to spiral galaxies in low- and the high-density environment as well the difference is larger for the intermediate-density environment. 

Our findings give rise to an interesting possibility. Does the intermediate-density environment is causing the bar formation in  S0 galaxies? If this is the case then these bars should have bluer colour and hence younger compare to spiral galaxies. The possibilities of minor mergers and tidal interactions are more in the intermediate-density environment which can form bars in S0 galaxies as shown in some numerical simulations and have been found observationally  \citep{Debattistaetal2002, Yangetal2009, Peiranietal2009}. At the same time, intermediate-density environment might lead to rejuvenation of star formation, particularly in bar region by facilitating the accretion of fresh gas from gas-rich satellites as suggested by HI observations \citep{Marino2011}.  

To summarise, we show that bars with blue colour are common in S0 galaxies compared to that in spiral galaxies. However, what makes these bars bluer is puzzling given that they are smaller in size and their bulges are old. Many of our barred S0 galaxies with bluer bar are in the intermediate-density environment, a possibility of a minor interaction is more and as stated earlier the bar formation can be triggered due to minor interactions.  More theoretical and simulation studies are required to explore the fate of the bars in S0 galaxies which can be the ideal laboratories for this investigation. Spatially resolved stellar populations of bars using IFU surveys such as MaNGA will provide useful insight and this can be the subject of future work.

%\section*{Acknowledgements}
\section*{Acknowledgements}
We thank the anonymous referee whose insightful comments have improved both the content and presentation of this paper. This research has made use of the NASA/IPAC Extragalactic Database (NED), which is operated by the Jet Propulsion Laboratory, California Institute of Technology (Caltech) under contract with NASA. We acknowledge the usage of the HyperLeda database (http://leda.univ-lyon1.fr). \\
Funding for the Sloan Digital Sky Survey IV has been provided by the Alfred P. Sloan Foundation, the U.S. Department of Energy Office of Science, and the Participating Institutions. SDSS-IV acknowledges
support and resources from the Center for High-Performance Computing at the University of Utah. The SDSS web site is www.sdss.org. \\
SDSS-IV is managed by the Astrophysical Research Consortium for the 
Participating Institutions of the SDSS Collaboration including the 
Brazilian Participation Group, the Carnegie Institution for Science, Carnegie Mellon University, the Chilean Participation Group, the French Participation Group, Harvard-Smithsonian Center for Astrophysics, 
Instituto de Astrof\'isica de Canarias, The Johns Hopkins University, Kavli Institute for the Physics and Mathematics of the Universe (IPMU) / 
University of Tokyo, the Korean Participation Group, Lawrence Berkeley National Laboratory, 
Leibniz Institut f\"ur Astrophysik Potsdam (AIP),  
Max-Planck-Institut f\"ur Astronomie (MPIA Heidelberg), 
Max-Planck-Institut f\"ur Astrophysik (MPA Garching), 
Max-Planck-Institut f\"ur Extraterrestrische Physik (MPE), 
National Astronomical Observatories of China, New Mexico State University, 
New York University, University of Notre Dame, 
Observat\'ario Nacional / MCTI, The Ohio State University, 
Pennsylvania State University, Shanghai Astronomical Observatory, 
United Kingdom Participation Group,
Universidad Nacional Aut\'onoma de M\'exico, University of Arizona, 
University of Colorado Boulder, University of Oxford, University of Portsmouth, 
University of Utah, University of Virginia, University of Washington, University of Wisconsin, 
Vanderbilt University, and Yale University.

\bibliographystyle{mnras}
\bibliography{sbarway.bib}

% Don't change these lines
\bsp	% typesetting comment
\label{lastpage}
\end{document}